\documentclass[12pt]{article}
\usepackage[top=2.5cm, bottom=1in, left=2.5cm, right=2.5cm]{geometry} 
\usepackage{textcomp}
\usepackage{color}         
\usepackage{graphics}      
\usepackage[pdftex]{graphicx}      
\usepackage{longtable}     
\usepackage{epsf}          
\usepackage{bm}            
\usepackage{asymptote}     
\usepackage{thumbpdf}
\usepackage[colorlinks=true]{hyperref} 
\usepackage{amsmath}
\usepackage{amssymb}
\usepackage{wrapfig}
\usepackage{enumerate}
\usepackage[toc,page]{appendix}
\usepackage{hyperref}
\usepackage{supertabular}
\usepackage{mdframed}
\usepackage[stable]{footmisc} 
\usepackage[utf8]{inputenc}
\usepackage{array}
\usepackage{subcaption}
\usepackage{capt-of}
\usepackage{multirow}
\usepackage{hhline}
\usepackage{arydshln}
\usepackage{tikz}

\usepackage{tocloft}
\newcommand{\ba}[1]{\begin{align} #1 \end{align} }

\newcommand{\bs}[1]{\begin{split} #1 \end{split} }

\def\tr{\text{Tr}}

\newcommand{\mc}[1]{\mathcal{ #1} }

\newcommand{\mbb}[1]{\mathbb{#1}}

\usepackage{titlesec}
\titleformat*{\section}{\large\bfseries}
\titleformat*{\subsection}{\normalsize\bfseries}
\titleformat*{\subsubsection}{\normalsize\it}

\usepackage{abstract}

\newcommand{\address}[1]{\vbox{\center\em#1}\vspace{9mm}}

\usepackage{lmodern}
\makeatletter
\patchcmd{\@maketitle}{\LARGE \@title}{\fontsize{25}{19.2}\selectfont\@title}{}{}
\makeatother

\usepackage[affil-it]{authblk} 
\usepackage{etoolbox}

\begin{document}
\title{\vspace{16mm}   4d SCFTs from negative-degree line bundles\vspace{4mm}}
\author{Emily Nardoni\footnote{enardoni@ucsd.edu}}
\date{\vspace{-18mm}}
\maketitle%
\address{Department of Physics, University of California, San
Diego, La Jolla, CA 92093 USA}


We construct  4d $\mc{N}=1$ quantum field theories by compactifying the (2,0) theories on a Riemann surface with genus $g$ and $n$ punctures, where the normal bundle decomposes into a sum of two line bundles with possibly negative degrees $p$ and $q$. Until recently, the only available field-theoretic constructions required the line bundle degrees to be nonnegative, although supergravity solutions were constructed in the literature for the zero-puncture case for all $p$ and $q$. Here, we provide field-theoretic constructions and computations of the central charges of 4d $\mc{N}=1$ SCFTs that are the IR limit of M5-branes wrapping a surface with general $p$ or $q$ negative, for general genus $g$ and number of maximal punctures $n$.

\vfill
November 2016 \vspace{3mm}

\newpage

{\hypersetup{linkcolor=black}
\tableofcontents
}

\section{Introduction} \label{sec:intro}        

\subsection{Background and motivations} 

One interesting way to obtain 4d quantum field theories is to compactify the 6d $\mc{N}=(2,0)$ superconformal field theories on a Riemann surface. As the (2,0) theories are the low-energy description of multiple M5-branes, the 4d theories that result from this procedure describe the low-energy limit of M5-branes that wrap the Riemann surface.  The 4d SCFTs that have such a description are known as theories of class $\mc{S}$.

This strategy for obtaining 4d theories has been used in an attempt to study and classify theories with $\mc{N}=2$ supersymmetry in \cite{Witten:1997sc,Gaiotto:2009we,Gaiotto:2009hg}. In \cite{Gaiotto:2009we}, it was argued that strongly coupled $\mc{N}=2$ SCFTs, denoted $T_N$ theories, arise as the low-energy effective description of $N$ coincident M5-branes wrapping a 3-punctured sphere. The punctures (i.e. codimension 2 defects) each introduce an $SU(N)$ flavor symmetry in the field theory. The $T_N$ theories can be combined by gauging diagonal subgroups of these global symmetries, which geometrically corresponds to gluing the punctures to form more general Riemann surfaces with arbitrary genus and punctures \cite{Gaiotto:2009we}. These constructions provided an understanding of the 4d SCFTs discussed in \cite{Maldacena:2000mw} which arise for the case of branes wrapping a surface with no punctures, and which are holographically dual  \cite{Gaiotto:2009gz} to some solutions of 11d supergravity referred to in the literature as Maldacena-Nu\~{n}ez solutions. The holographic duals of constructions for the more general system with punctures were considered in \cite{Gaiotto:2009gz}.

One can field-theoretically construct $\mc{N}=1$ theories of class $\mc{S}$---including the duals of the $\mc{N}=1$ Maldacena-Nu\~{n}ez solutions---via deformations of the $T_N$ theories, as discussed in \cite{Benini:2009mz}. This program was continued in \cite{Bah:2011je} with constructions in which the $T_N$ theories are coupled with $\mc{N}=1$ vector multiplets. In the present work, we will be interested in generalizing an infinite class of 4d $\mc{N}=1$ SCFTs that were introduced in \cite{Bah:2011vv,Bah:2012dg}. There, supergravity solutions that naturally generalize the Maldacena-Nu\~{n}ez solutions (and which we'll refer to as BBBW solutions for authors Bah, Beem, Bobev, Wecht) were constructed and shown to be holographically dual to field-theoretic constructions that involve $\mc{N}=1$ gluing of $T_N$ building blocks to create a surface with no punctures. 

We here consider M5-branes wrapping a genus $g$ Riemann surface with $n$ maximal punctures $\mc{C}_{g,n}$, where the surface is embedded in a Calabi-Yau 3-fold\footnote{Eight supercharges are preserved when the ambient geometry in which the Riemann surface is embedded in a Calabi-Yau 2-fold, while only four supercharges are preserved when the geometry is a Calabi-Yau 3-fold. For further discussion on the general structure of how $\mc{N}=1$ supersymmetry is preserved in the BBBW theories, see \cite{Bah:2012dg}.}. The normal bundle decomposes into a sum of two line bundles as
	\ba{ \bs{
	\begin{tikzpicture}
	\node at (0,0) {$\mbb{C}^2$};
	\draw[->] (0.4,0) -- (1,0);
	\node at (1.8,0) {$\mc{L}_1\oplus \mc{L}_2$};
	\draw[->] (1.8,-.3) -- (1.8,-1);
	\node at (1.8,-1.3) {$\mc{C}_{g,n}$};  \label{eq:cygeometry} 
	\end{tikzpicture} }}
The Calabi-Yau condition requires that the degrees of the line bundles $p$ and $q$ satisfy 
	\ba{
p\equiv c_1(\mc{L}_1),\qquad  q\equiv c_1(\mc{L}_2),\qquad p+q=-\chi(C_{g,n})=2g-2+n. \label{eq:generaln}
	}
The BBBW supergravity solutions are valid for all $p$ and $q$; however, there is only an explicit field theory construction for the case of $p$ and $q$ nonnegative\footnote{See note added at the end of the Introduction; similar constructions were recently considered in \cite{Fazzi:2016eec}.}. Our main goal in the present work is to understand such a construction for the case where one of the line bundle degrees is negative. Our construction requires a more general building block than the $T_N$ theory. The necessary ingredient was provided in  \cite{Agarwal:2015vla}, which introduced a generalization of the $T_N$ theory denoted $T_N^{(m)}$, for $m$ a positive integer (and whose features we will review in Section \ref{sec:tnm}). The field-theoretic constructions in \cite{Agarwal:2015vla} utilizing $T_N^{(m)}$ building blocks provided the first generalized quiver field theories with $p,q<0$.

\subsection{Overview and summary of results} 

In the present work, we explicitly construct 4d $\mc{N}=1$ field theories that result from compactifying the (2,0) theories on a surface with negative $p$ or $q$, thereby providing field theoretic constructions for the duals of the BBBW gravity solutions obtained by gluing $T_N^{(m)}$ building blocks. We further generalize to the case of M5-branes wrapped on Riemann surfaces with maximal punctures, yielding formulae for the (trial) central charges of the resulting SCFTs that depend only on geometric data.

The organization of this paper is as follows. In Section \ref{sec:second}, we provide the field-theoretic construction of the 4d $\mc{N}=1$ SCFTs that are dual to the BBBW gravity solutions for negative $p$ or $q$. We begin by reviewing the definition of the $T_N^{(m)}$ theories as formulated in \cite{Agarwal:2015vla}. Then, we glue together $(2g-2)$ copies of the $T_N^{(m)}$ theories, yielding a genus $g>1$ surface with no punctures and possibly negative $p$ or $q$, thus providing an inherently field-theoretic construction of 4d $\mc{N}=1$ field theories that arise as the IR limit of M5-branes wrapped on a surface with negative normal bundle degrees. We compute the central charges and operator dimensions for these theories, and find that they match precisely onto the BBBW formulae. 

In Section \ref{sec:spheresec}, we study the genus zero case, which requires closing punctures on chains of $T_N^{(m)}$ theories. We consider the simplest case of a single $T_N^{(m)}$ theory whose $SU(N)$ flavor groups are Higgsed, which can yield theories with twist $|z|\geq 2$, and find that the trial central charges match onto the BBBW results. There will be corrections to these values from operators whose R-charges were shifted in the Higgsing procedure. We comment on these corrections, but leave their complete analysis to future work.

Having reviewed the machinery to close punctures in Section  \ref{sec:spheresec}, in Section \ref{sec:cgn} we construct theories of genus $g$ and $n$ maximal punctures from $T_N^{(m_i)}$ building blocks. With these general constructions, we are able to compute the central charges for the torus as well. We conclude with a brief discussion of future directions. Appendices review our conventions, relevant BBBW formulae we reference throughout, and list the 't Hooft anomalies needed in our constructions.

{\bf Note added:} While this work was in the final stages of preparation, \cite{Fazzi:2016eec} appeared, which has some overlap with the present work. 


\section{Constructing the BBBW duals from  $T_N^{(m)}$ building blocks} \label{sec:second}

\subsection{$T_N^{(m)}$ review} \label{sec:tnm} 

\begin{figure}[t]
\centering
\includegraphics[width=0.9\textwidth]{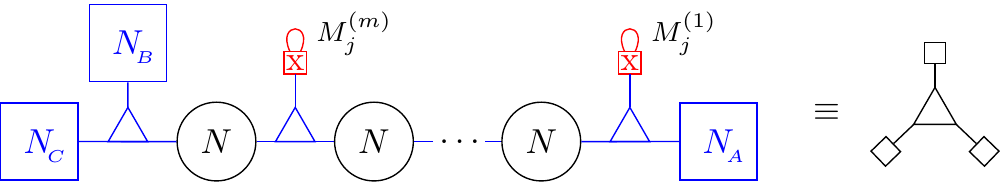}
\caption{A UV generalized quiver description for the $T_N^{(m)}$ theories. The blue color of the $SU(N)_{A,B,C}$ flavor groups corresponds to punctures with sign $\sigma_{A,B,C}=+1$. The red boxes correspond to closed $\sigma=-1$ punctures, i.e. red-colored $SU(N)$ flavor groups that were Higgsed by giving vevs to adjoint operators $M^{(i)}$. The singlets $M_j^{(i)}$ are the leftover components of the fluctuations of the $M^{(i)}$ about the vevs. The diagram to the right of the quiver is used in later figures as a shorthand. \label{fig:tnmquiver}}
\end{figure}

The $T_N^{(m)}$ theories constructed in  \cite{Agarwal:2015vla} are strongly coupled 4d $\mc{N}=2$ SCFTs. They have a natural description as being of class $\mc{S}$, arising as the low-energy limit of wrapping $N$ M5-branes (the (2,0) theories of type $A_{N-1}$) on a sphere with three punctures, $\mc{C}_{0,3}$. The sphere is embedded in a Calabi-Yau 3-fold, which decomposes into a sum of two line bundles as in  \eqref{eq:cygeometry}. For our purposes, the novelty of this construction is that the degrees $p$ and $q$ of the line bundles are allowed to be negative, parameterized by a nonnegative integer $m$:
	\ba{
	T_N^{(m)}:\quad p=m+1,\quad q=-m. \label{eq:tnmdegrees}
	} 
For $m=0$, this construction reduces to the $T_N$ theory.  

These theories preserve a $U(1)_+\times U(1)_-$ global symmetry which derives from the parent $\mc{N}=(2,0)$ $SO(5)_R$ symmetry, as well as an $SU(N)^3$ global symmetry associated to the three punctures\footnote{This is taking the punctures to be maximal; to construct building blocks with generic three punctures whose flavor symmetries are non-maximal, one can use results in \cite{Gadde:2013fma}, \cite{Agarwal:2014rua}, \cite{Chacaltana:2010ks}, \cite{Chacaltana:2012zy}.} which we denote $SU(N)_{A}\times SU(N)_B\times SU(N)_C$. Each puncture is labeled by a sign $\sigma_{A,B,C}=\pm1$; in the notation of \cite{Agarwal:2015vla}, +1 is blue-colored and -1 is red-colored, and in diagrams we'll take $+1$ to be unshaded and $-1$ to be shaded. Denoting the generators of $U(1)_+\times U(1)_-$ by $J_\pm$, the exact superconformal R-symmetry is given by the linear combination (see Appendix \ref{sec:conventions} for more on our conventions)
	\ba{
	R_{\mc{N}=1}(\epsilon) = \frac{1-\epsilon}{2} J_+ + \frac{1+\epsilon}{2}J_-. \label{eq:rcharge}
	}	
The 't Hooft anomalies for the $T_N^{(m)}$ theories are given in Appendix \ref{sec:anomalies}, in equation \eqref{eq:tnmanomalies}, and the chiral operators of the $T_N^{(m)}$ theories are listed in Table \ref{tab:tnmatter}. The $\phi_i$ are adjoint chiral multiplets; the $\mu_{j}^{(i)}$ and $M_j^{(i)}$ are singlets; the $\mu_i$ are moment-map operators (i.e., chiral operators at the bottom of would-be $\mc{N}=2$ current multiplets) of the $m$ symmetry groups $SU(N)_i$ that are gauged in the construction of the theories; the $\mu_{A,B,C}$ are moment-map operators of the leftover $SU(N)_{A,B,C}$ flavor symmetries; and $Q (\widetilde{Q})$ are (anti)trifundamentals of the $SU(N)_A\times SU(N)_B\times SU(N)_C$ flavor symmetry. Some of these operators are coupled in a superpotential,
	\ba{
W = \sum_{i=1}^m \tr \phi_i (\hat{\mu}_i - \widetilde{\hat{\mu}}_i) + \sum_{i=1}^m \sum_{j=1}^{N-1} \mu_{j}^{(i)} M_j^{(i)}. \label{eq:sup}
	}
The origin of this superpotential comes from the construction of the $T_N^{(m)}$ theories from $m+1$ copies of the $T_N$ theory; see \cite{Agarwal:2015vla} for more details. We summarize the resulting UV generalized quiver description for the $T_N^{(m)}$ theories in Figure \ref{fig:tnmquiver}.

The superpotential \eqref{eq:sup} yields chiral ring relations for the chiral operators. For example, while naively one might worry that the singlets $\mu_j^{(i)}$ could violate the unitarity bound due to their negative $J_-$ charge, the F-terms for the $M_j^{(i)}$ imply that the singlets $\mu_j^{(i)}$ are in fact trivial in the ring. As checked in \cite{Agarwal:2015vla}, none of the gauge-invariant chiral operators that are nontrivial in the ring decouple\footnote{If the dimension of a chiral operator $\mc{O}$ appears to violate the unitarity bound $R(\mc{O}) <2/3$ (the R-charge of a chiral operator is proportional to its dimension in theories with four supercharges), then $\mc{O}$ is in fact free, and an accidental $U(1)$ symmetry acts on $\mc{O}$. One must account for the decoupling of these free operators, e.g. in computations of $a$ and $c$ \cite{Kutasov:2003iy}.}. 

	\begin{table}[t!]
	\centering 
	\begin{tabular}{|c|c|c|c|c|c|} 
	\hline
	 & $SU(N)_i$ & $SU(N)_A$ & $SU(N)_B$ & $SU(N)_C$ & $(J_+,J_-)$ \\ \hline  
	$Q$ &  & $\Box$ & $\Box$ & $\Box$ & $(N-1,0)$ \\
		$\widetilde{Q}$ &  & $\bar{\Box}$ & $\bar{\Box}$ & $\bar{\Box}$ & $(N-1,0)$ \\
		$\phi_i\ (1\leq i\leq m)$ & adj & & & & $(0,2)$ \\
	$\mu_i\ (1\leq i\leq m)$ & adj & & & & $(2,0)$ \\
	$\mu_A$ &  &adj  & & & $(2,0)$ \\
	$\mu_B$ &  &  &adj & & $(2,0)$ \\
	$\mu_C$ &  & &  & adj& $(2,0)$ \\
	$\mu_j^{(i)}\ (1\leq j\leq N-1)$ &  & &  & & $(2,-2j)$ \\
	$M_j^{(i)}\ (1\leq j\leq N-1)$ &  & &  & & $(0,2j+2)$ \\ [3pt]
	\hline
	\end{tabular} \caption{Operators of the (unshaded) $T_N^{(m)}$ theories. \label{tab:tnmatter}}
	\end{table}

\subsection{Gluing procedure}

The gluing procedure corresponds to decomposing the geometry into pairs of pants, where each can be associated with $T_N^{(m)}$ theories, and gauging subgroups of the flavor symmetries associated to the punctures. We will label the $i$th block by $T_N^{(m_i)}$, where the $m_i$ are in general different. 

Before we do the general case, let us first illustrate the procedure of gluing two $T_N^{(m_i)},\ i=1,2$ theories with either an $\mc{N}=1$ or $\mc{N}=2$ vector multiplet, as in \cite{Agarwal:2015vla}. Label the degrees of the blocks as
	\ba{
	T_N^{(m_i)}:\quad (p_i,q_i)= \left\{ \begin{array}{cl} (m_i+1,\ -m_i) & \quad \sigma_i=+1 \\ (-m_i,\ m_i+1)& \quad\sigma_i=-1 \end{array} \right. \label{eq:degreestnm}
	}
since $p_i$ and $q_i$ switch roles for an unshaded versus shaded block. We gauge an $SU(N)$ flavor symmetry of the two $T_N^{(m_i)}$ theories, leading to a superpotential for the moment-map operators of the gauged block,
	\ba{
	W=\tr \mu^+ \mu^-.
	}
For instance, $\mu^+$ could be chosen to derive from $\mu_A$ in Table \ref{tab:tnmatter} for one block, and $\mu^-$ to come from $\mu_A$ for the other block. In order to write a superpotential of this form, when gluing with an $\mc{N}=1$ vector the $(J_+,J_-)$ charge assignment of one of the $T_N^{(m_i)}$ blocks must be flipped such that $\mu^-$ has $J_\pm$ charges given by $(J_+,J_-)=(0,2)$. In general, two blocks of the same color/shading should be glued by an $\mc{N}=2$ vector, while two blocks of differing colors/shadings should be glued by an $\mc{N}=1$ vector.

The result of this procedure is a four-punctured sphere $\mc{C}_{0,4}$, where the total degrees $p$ and $q$ of the embedding space satisfy $p+q=2g-2+n=2$.

\subsection{Construction of $\mc{C}_{g>1,n=0}$ and computation of $a$ and $c$}
\label{sec:b3wanalogue}

\begin{figure}[t]
\centering
\includegraphics[width=0.6\textwidth]{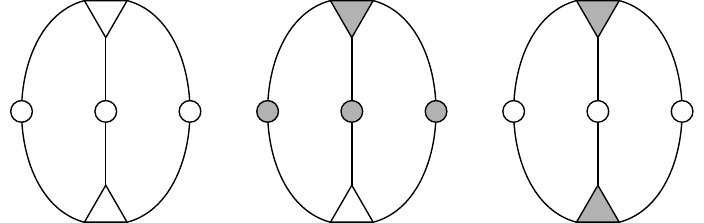}
\caption{Options for gluing 2 $T_N^{(m_i)}$ blocks to form a genus $2$ surface. Shaded triangles correspond to $T_N^{(m_i)}$ theories with $\sigma_i=-1$, while unshaded triangles have $\sigma_i=+1$. Shaded circular nodes correspond to $\mc{N}=1$ vector multiplets, while unshaded circular nodes correspond to $\mc{N}=2$ vector multiplets. Blocks of differing colors should be glued by an $\mc{N}=1$ vector. \label{fig:genus2ex}}
\end{figure}

Here, we glue together $(2g-2)$ copies of the $T_N^{(m_i)}$ theories, $i=1,...,2g-2$, yielding a genus $g>1$ surface  $\mc{C}_{g,0}^{(p,q)}$ with no punctures and possibly negative degrees $p$ and $q$. (The sphere and torus cases are constructed separately in later sections.) We consider the general case of $\ell_1$ (shaded) blocks with $\sigma_i=-1$, and $\ell_2$ (unshaded) blocks with $\sigma_i=+1$, glued together with $n_1$ (shaded) $\mc{N}=1$ vector multiplets and $n_2\ \mc{N}=2$ (unshaded) vector multiplets. Given the geometries we wish to construct, these parameters satisfy
	\ba{
	\ell_1+\ell_2=2(g-1),\qquad n_1+n_2=3(g-1). \label{eq:parameters}
	}
Label the degrees of the blocks as in \eqref{eq:degreestnm}. Then, the total degrees $p$ and $q$ of the space in which the genus $g$ surface is embedded are given by
	\ba{
p =  \sum_{\{\sigma_i=+1\}} m_i - \sum_{\{\sigma_i=-1\}} m_i + \ell_2,\qquad p+q=2g-2.
	}
The sum over $\{\sigma_i=+1\}$ runs over the $\ell_2$ unshaded nodes, while the sum over $\sigma_i=-1$ runs over the $\ell_1$ shaded nodes. As an example, the options for forming a genus 2 surface in this manner are shown in Figure \ref{fig:genus2ex}.

We now compute the central charges $a$ and $c$ for these configurations. For a general 4d $\mc{N}=1$ SCFT, the central charges $a$ and $c$ are determined by the 't Hooft anomalies \cite{Anselmi:1997am},
	\ba{
	a=\frac{3}{32}\left( 3 \tr R^3 - \tr R \right),\quad c = \frac{1}{32} \left(9\tr R^3 - 5 \tr R\right). \label{eq:centralcharge}
	}
For quivers made from $T_N^{(m)}$ building blocks, in the absence of accidental symmetries,  the $\mc{N}=1$ superconformal R-symmetry $R=R(\epsilon)$ takes the form (see Appendix \ref{sec:conventions} for conventions)
	\ba{
	R(\epsilon)=\frac{1}{2}(1-\epsilon)J_+ + \frac{1}{2}(1+\epsilon)J_- . \label{eq:rcharge2}
	}
Then, the exact superconformal R-symmetry at an IR fixed point is determined by $a$-maximization with respect to $\epsilon$ \cite{Intriligator:2003jj}. 

The contributions to $a$ and $c$ of the various components of our constructions can be computed using the 't Hooft anomalies given in Appendix \ref{sec:anomalies}, substituted into equations \eqref{eq:centralcharge} and \eqref{eq:rcharge2}. The contribution of the $i$'th $T_N^{(m_i)}$ block is given by \cite{Agarwal:2015vla}
		\ba{ \bs{
a_{T_N^{(m_i)}}(\epsilon) = &\frac{3}{64} (N-1)(1-\epsilon) \left(3N^2(1+\epsilon)^2 - 3N (2\epsilon^2 + \epsilon + 1) - 2(3\epsilon^2 + 3\epsilon + 2) \right)  \\
&-m_i\frac{3}{32}  \epsilon \left(3N^3 (\epsilon^2-1) -3\epsilon^2 + 2N +1\right)\\
\equiv &\  \mc{A}_0(\epsilon) + m_i \mc{A}_1(\epsilon).} \label{eq:pieces}
	}
For convenience, we've defined $\mc{A}_0(\epsilon)$ as the piece of $a_{T_N^{(m_i)}}(\epsilon)$ that's independent of $m_i$, and  $\mc{A}_1(\epsilon)$ as the piece proportional to $m_i$. Our convention is that these formulae as written correspond to an unshaded ($\sigma_i=+1$) block, while taking $\epsilon\to - \epsilon$ (equivalently, swapping $J_+$ and $J_-$) yields the formulae for a shaded block. The contributions of an $\mc{N}=2$ and $\mc{N}=1$ vector multiplet are
	\ba{
	a_{\mc{N}=2}(\epsilon) &= \frac{3}{32}(N^2-1) \epsilon (3 \epsilon^2-1) + \frac{6}{32}(N^2-1) \label{eq:neq2} \\
	a_{\mc{N}=1} &= \frac{6}{32} (N^2-1). \label{eq:neq1}
	}
Here, the convention for $a_{\mc{N}=2}(\epsilon)$ is that as written we're gluing two unshaded flavor groups\footnote{Our conventions appropriately account for this, e.g. by not including an absolute value in the definition of $n_2$ in \eqref{eq:parameters}.}; gluing two shaded flavor groups with an $\mc{N}=2$ vector corresponds to taking $\epsilon\to-\epsilon$. Then, $a(\epsilon)$ for this class of theories is given by a sum over these pieces,
	\ba{\bs{
	a(\epsilon) =& \ell_2 \mc{A}_0 (\epsilon) + \mc{A}_1(\epsilon) \sum_{\{\sigma_i=+1\}} m_i + \ell_1 \mc{A}_0(-\epsilon) + \mc{A}_1 (-\epsilon) \sum_{\{\sigma_i=-1\}} m_i \\
	&+ \frac{3}{2} (\ell_2-\ell_1)a_{\mc{N}=2}(\epsilon) + 3\ell_1 a_{\mc{N}=1}. }\label{eq:aquiver}
	}	
Maximizing with respect to $\epsilon$ yields
	\ba{ 
	\epsilon = \frac{N+N^2-\sqrt{z^2+N(1+N) \big( N(1+N)+z^2(4+3N(1+N)) \big)}}{3(1+N+N^2)z}, \label{eq:eps0}
	}
where we've written the answer in terms of the twist parameter $z$, 
	\ba{
	z=\frac{p-q}{p+q},\qquad p=(g-1)(1+z),\qquad q=(g-1)(1-z). \label{eq:zdef}
	}
$\epsilon$ in \eqref{eq:eps0} matches the value computed in \cite{Bah:2012dg}, as expected. The argument of the square root is always positive for the valid ranges of the parameters, $N\geq 2,\ g\geq 2$. $\epsilon$ is singular only for $q=p=g-1$, and re-maximizing with respect to $\epsilon$ for this special point yields $\epsilon=0$.

Substituting $\epsilon$ into \eqref{eq:aquiver}, and performing the similar computation for $c$, we find
	\ba{ \bs{
	a=\frac{(N-1)(g-1)}{48 (1+N+N^2)z^2} \bigg[ &-N^3(1+N)^3 +3z^3(1+N+N^2) (3+N(1+N) (7+3N(1+N))) \\
	& +\big(z^2+N(1+N)( N(1+N) + (4+3N(1+N))z^2)\big)^{3/2} \bigg].
	}\label{eq:a0} \\
		\bs{
	c=\frac{(N-1)(g-1)}{48(1+N+N^2)^2z^2} \bigg[& -N^3(1+N)^3 + z^2(1+N+N^2)(6+N(1+N)(17+9N(1+N))) \\
	& +(N^2(1+N)^2 + z^2(1+N+N^2)(2+3N(1+N)) \\
	&\ \cdot \sqrt{z^2 + N (1+N) \big(N (1+N) + z^2(4+3N(1+N))\big)}. \bigg]
	} \label{eq:c0}
	}
We plot $a$ as a function of $z$ for various values of $N$ in Figure \ref{fig:apic0}. These results precisely match the values of $a$ and $c$ that were computed by BBBW in \cite{Bah:2012dg} via integrating the anomaly eight-form of the M5-brane theory over $\mc{C}_{g,0}$. We emphasize that we've found this same result with a field theoretic construction. It is worth highlighting that it is nontrivial that the dependence on the $m_i$ in \eqref{eq:aquiver} cancels to yield central charges \eqref{eq:a0} and \eqref{eq:c0} that depend only on the topological data $z$ and $g$ (and choice of $N$).

\begin{figure}[t]
\centering
\includegraphics[width=0.7\textwidth]{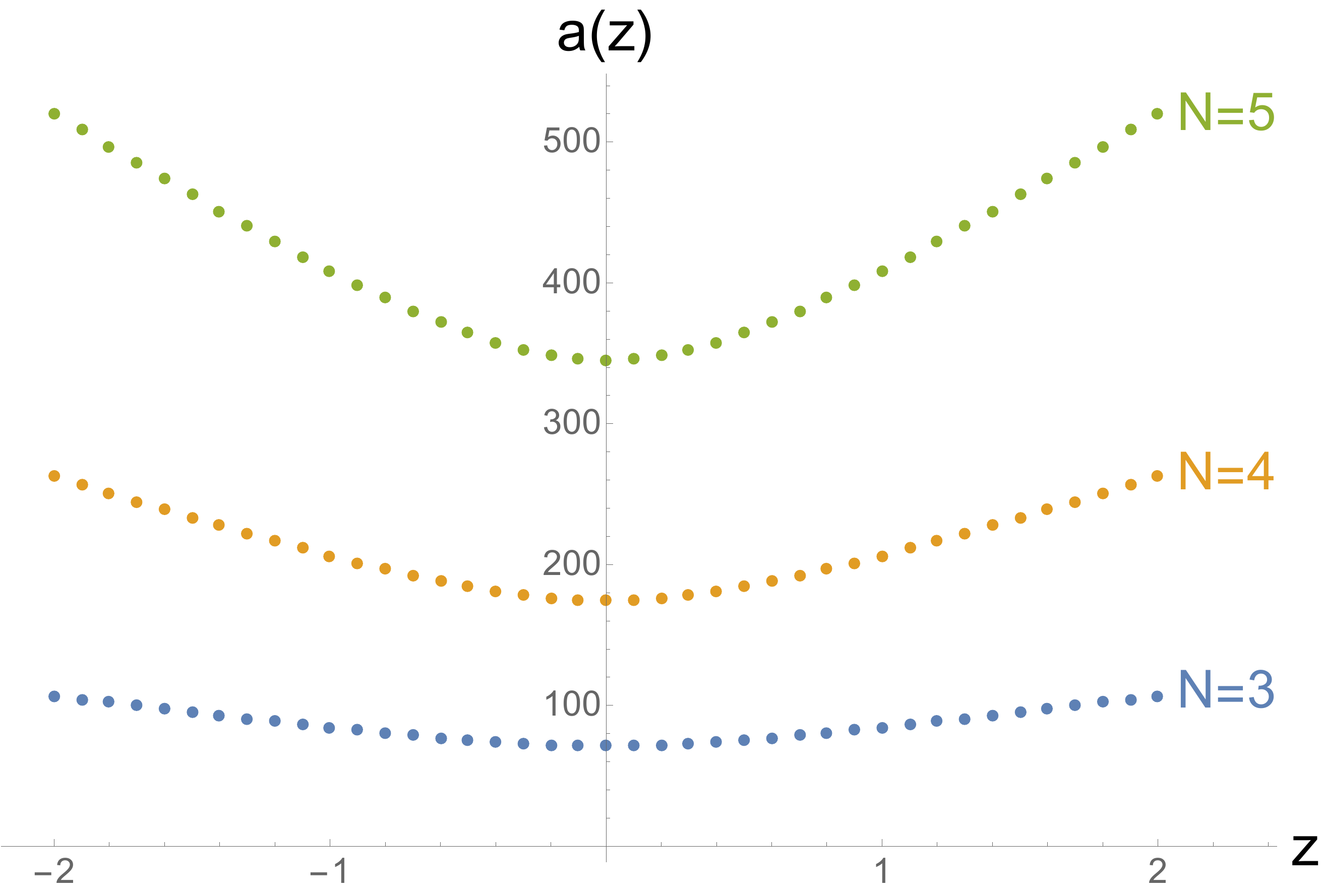}
\caption{$a$ as a function of the twist $z$ for quivers constructed from $2g-2$ glued $T_N^{(m_i)}$ blocks, plotted for $g=11$ and various $N$. \label{fig:apic110}}
\end{figure}

\subsection{Operator dimensions and large-$N$} \label{sec:opdims}

With $\epsilon$ fixed in \eqref{eq:eps0}, the dimensions of chiral operators can be determined by $\Delta (\mc{O})=\frac{3}{2}R(\mc{O})$, using $R(\epsilon)$ defined in \eqref{eq:rcharge}. The dimensions of operators are then given by
	\ba{ \renewcommand\arraystretch{1.3} \begin{array}{ll}
 \Delta[\mu]=\frac{3}{2} (1- \epsilon), &\  \Delta[\phi] = \frac{3}{2} (1+ \epsilon)\\
 \Delta[Q] = \frac{3}{4} (N-1)(1-\epsilon), &  \Delta[u_k]=\frac{3}{2}(1+\epsilon) k,\ \ k=3,\dots, N\ \\
\Delta[M_j]=\frac{3}{2} (1+\epsilon)(1+j),\ \  j=1,\dots, N-1. &  \end{array} \label{eq:opdims}
	}
Again, we use a convention where taking $\epsilon\to - \epsilon$ yields the R-charge of the operator corresponding to a shaded node; e.g.  $\Delta[\phi] (\epsilon)$ corresponds to the adjoint chiral field in the $\mc{N}=2$ vector multiplet of an unshaded node, while $\Delta[\phi ](-\epsilon)$ corresponds to the adjoint chiral of a shaded node. The $\mu$ are the various moment map operators of the $T_N^{(m_i)}$ blocks, detailed in Table \ref{tab:tnmatter}.

We can construct gauge-invariant operators out of $Q$ and $\widetilde{Q}$ that correspond to M2-brane operators wrapping the surface $\mc{C}_{g,0}$, as described in \cite{Bah:2012dg}. Schematically, these are
	\ba{
\mc{O}_{M2} = \prod_{i=1}^{2g-2}Q_i,\qquad \widetilde{\mc{O}}_{M2} = \prod_{i=1}^{2g-2}\widetilde{Q}_i
	} 
From \eqref{eq:opdims}, the dimensions of these operators are
	\ba{ \bs{
\Delta [\mc{O}_{M2}]=\Delta [\widetilde{\mc{O}}_{M2}] 
&= \frac{3}{4} (N-1)  \left[ (2g-2) +\epsilon (\ell_1-\ell_2) \right]
	}}
for $\epsilon$ given in \eqref{eq:eps0}, and where $0\leq |\ell_1-\ell_2|\leq 2g-2$. None of these operators decouple.

$|\epsilon|$ in \eqref{eq:eps0} ranges from 0 at $z=0$, to $\frac{1}{\sqrt{3}}$ at large $z$ and $N$. In particular, the new range of $\epsilon$ accessible for negative $p$ and $q$ versus the previously studied case of $p,q\geq 0$ \cite{Bah:2012dg} is the range $\frac{1}{3} \leq |\epsilon| \leq \frac{1}{\sqrt{3}}$. In this range of $\epsilon$, no operators violate the unitarity bound; thus, assuming no accidental IR symmetries, $a$ and $c$ are given by \eqref{eq:a0} and \eqref{eq:c0}. We note that $a$ and $c$ given in \eqref{eq:a0}, \eqref{eq:c0} are always positive and nonimaginary, and always (for $g>1$) satisfy the Hofman-Maldacena bounds for $\mc{N}=1$ SCFTs \cite{Hofman:2008ar},
	\ba{
	\frac{1}{2}\leq \frac{a}{c} \leq \frac{3}{2}.
	}
At large $N$, the leading-order term of $a=c$ in \eqref{eq:a0} scales as $N^3$, and is given by
	\ba{
 a_{\text{large-}N} = \frac{(1-g) (1-9z^2-(1+3z^2)^{3/2})}{	48z^2} N^3. \label{eq:alargen}
	}
This reproduces equation (2.22) in \cite{Bah:2012dg}.

\section{$\mc{C}_{g=0,n=0}$ from the higgsed $T_N^{(m)}$}
\label{sec:spheresec}

\subsection{Constructing higgsed $T_N^{(m)}$ theories}

In this section, we review the procedure of closing the three maximal punctures of the $T_N^{(m)}$ theory. The Higgsing procedure we review below was detailed in  \cite{Agarwal:2015vla}, and also utilized in \cite{Maruyoshi:2016aim} (where the $|z|=2$ case was first studied) and \cite{Fazzi:2016eec}. 

\begin{figure}[t]
\centering
\includegraphics[width=1\textwidth]{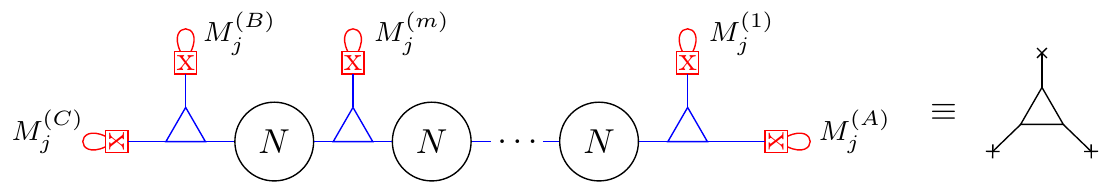}
\caption{UV quiver for the $T_N^{(m)}$ theory with $\sigma_i=+1$, Higgsing the $SU(N)$ flavor nodes. The rightmost figure depicts a shorthand we use throughout, for reference. \label{fig:sphere}}
\end{figure}

First, switch the color of the punctures to be opposite the color of the background $T_N^{(m)}$ (i.e. flip the colors of the $SU(N)_{A,B,C}$ flavor groups in Figure \ref{fig:tnmquiver} from blue to red), and couple in three extra chiral fields $M_{A,B,C}$ that transform as the adjoint of the $SU(N)_{A,B,C}$ flavor groups, respectively. The superpotential contains terms that couple these adjoints to the moment map operators $\mu_{A,B,C}$,
	\ba{
W\supset \mu_A M_A + \mu_B M_B + \mu_C M_C.
	}
Next, Higgs each of the flavor groups $SU(N)_{A,B,C}$ via a nilpotent vev\footnote{In general, one could consider a nilpotent vev corresponding to an $SU(2)$ embedding $\rho: SU(2)\to SU(N)$ labeled by a partition of $N$, with the residual flavor symmetry given by the commutant of the embedding---see \cite{Agarwal:2014rua} for more details. Here, we consider only the principal embedding.} 
	\ba{
	\langle M_{A}\rangle = \langle M_{B}\rangle =\langle M_{C}\rangle = \left( \begin{array}{ccccc} 0 & 1 & & &  \\ & 0 & 1 & &  \\ & & \ddots & \ddots &  \\ & & &  0 & 1 \\ & & &  & 0 \end{array} \right).
	}
A vev of this form corresponds to the principle embedding of $SU(2)$ into $SU(N)$, breaking each flavor group down from $SU(N)_{A,B,C}\to U(1)$. The adjoint representation of $SU(N)$ decomposes into a sum of spin-$j$ irreducible representations of $SU(2)$, such that the components of $M_{A,B,C}$ corresponding to fluctuations from the vev are labeled by the spin $j$ and the $\sigma_3$-eigenvalue $m=-j,\dots, j$ (e.g. see relevant discussion in  \cite{Agarwal:2014rua}  and \cite{Agarwal:2015vla}). The only components that don't decouple are those with $m=-j$, which we'll denote as $M_{j}^{(A,B,C)}$ below. After decoupling operators, the remaining superpotential is
 	\ba{
W\supset \sum_{j=1}^{N-1} \left(M^{(A)}_j \mu^{(A)}_j +  M^{(B)}_j \mu^{(B)}_j+  M^{(C)}_j \mu^{(C)}_j\right).
	}
The UV quiver is depicted in \ref{fig:sphere}.

The Higgsing shifts the R-charges 
	\ba{
	J_+\to J_+,\quad J_- \to J_-- \sum_{i=A,B,C} 2m^{(i)} \label{eq:rchargeshift}
	}
for $m^{(i)}$ the weights of the $SU(2)$ representations. The resulting R-charges of operators are given in Table \ref{tab:tnmhiggsedmatter}. Note that the trifundamental $Q$'s have decomposed into $N^3$ singlets, with R-charges shifted due to \eqref{eq:rchargeshift}.
	\begin{table}[h!]
	\centering 
	\begin{tabular}{|cc|c|c|} 
	\hline
	 & & $SU(N)$  & $(J_+,J_-)$ \\ \hline  
	$Q^{(s)(t)(u)},\widetilde{Q}^{(s)(t)(u)}$ & $-\frac{(N-1)}{2}\leq \{s,t,u\} \leq \frac{N-1}{2}$ &  & $\big(N-1,-2(s+t+u)\big)$ \\
		$\phi_i$ & $(i\leq 1\leq m)$ & adj  & $(0,2)$ \\
	$\mu_i$ & $(i\leq 1\leq m)$ & adj  & $(2,0)$ \\
	$\mu^{(A,B,C)}_j$ & $(1\leq j\leq N-1)$ &   & $(2,-2j)$ \\
	$\mu_j^{(i)}$ & $(1\leq j\leq N-1)$ &  & $(2,-2j)$ \\
	$M_j^{(i)}$ & $(1\leq j\leq N-1)$ &  & $(0,2j+2)$ \\ 
		$M_j^{(A,B,C)}$ & $(1\leq j\leq N-1)$ &  & $(0,2j+2)$ \\ [3pt]
	\hline
	\end{tabular} \caption{Operators in the Higgsed $T_N$ theory. \label{tab:tnmhiggsedmatter}}
	\end{table}
	
Geometrically, closing the punctures reduces the degrees of the normal bundle; starting with $p=m+1,\ q=-m$ as in \eqref{eq:tnmdegrees}, we flow to a theory with
	\ba{\bs{
p = m+1,\qquad q=-m-3&,\qquad m\geq 0 \\
\Rightarrow  z=-m-2&.
}\label{eq:pqsphere}
	}
Note that this construction only yields 4d theories with $|z|\geq 2$, since starting with a $T_N^m$ theory with $\sigma=\pm1$ results in $z=\mp (m+2)$ for $m\geq 0$.

\subsection{Computation of $a_{trial}$ and $c_{trial}$}

Next, we compute $a$ and $c$ for the Higgsed $T_N^{(m)}$ theories, assuming a flow to an IR fixed point. The central charges can be computed from the 't Hooft anomalies for the $T_N^{(m)}$ theory given in Appendix \ref{sec:anomalies}, adding in the contribution from Higgsing the $SU(N)_{A,B,C}$ symmetries of the three punctures given in \eqref{eq:higgsedanom}. The contribution to $a$ from closing a single puncture can be expressed as
	\ba{
a_{\langle M \rangle}(\epsilon) = \frac{3}{64} \left( 2\epsilon - 6 \epsilon^3 + 3N^3(\epsilon-1)(\epsilon+1)^2 + N(1+(2-3\epsilon)\epsilon) + N^2 (2-\epsilon+3\epsilon^3) \right). \label{eq:apunct} 	
	}
With these ingredients, we find that $\epsilon$ is given by
	\ba{
	\epsilon= \frac{N+N^2 + \sqrt{z^2 + N(1+N)( N(1+N) + z^2(4+3N(1+N)))}}{3(1+N+N^2)z}
	}
and the central charges $a$ and $c$ are given by
	\ba{\bs{
a_{trial}=\frac{1}{48 (1+N+N^2)^2z^2}\bigg[&(1+N)^3 -z^2 (1+N+N^2)(9+3N(1+N)(7+3N(1+N)))\\
& \cdot \big(z^2+N(1+N)( N(1+N) + z^2(4+3N(1+N)))\big)^{3/2} \bigg], \\
c_{trial}=\frac{(N-1)}{48(1+N+N^2)^2 z^2} \Bigg[&N^3(1+N)^3 + \big(z^2 + N(1+N)(N(1+N) + z^2(4+3N(1+N)) )\big)^{3/2} \\
& - z^2 (1+N+N^2) \bigg[6+N(1+N) (17+9N(1+N)) \\
& - \sqrt{z^2 + N(1+N)(N (1+N) + z^2(4+3N (1+N)))} \bigg]\Bigg].
} \label{eq:asphere}
	}
These match the BBBW results, given in \eqref{eq:abbbw} of Appendix \ref{sec:bbbwapp} with $\kappa=1$ and $g=0$. However, as we discuss in the next section, this is not the whole story, and there will be field theory corrections from operators in the theory hitting the unitarity bound. For this reason, we explicitly include the label $a_{trial},\ c_{trial}$.

\subsection{Comments on ruling out $g=0$ SCFTs}

For the Higgsed $T_N^{(m)}$ theory, the chiral operators are summarized in Table \ref{tab:tnmhiggsedmatter}. $|\epsilon|$ runs from $\frac{1}{3}\sqrt{\frac{19}{7}}$ for $N=2$ and large-$z$, to $\frac{1}{6} (1+\sqrt{13} )$ for $z=2$ and large-$N$. For instance, the singlets $Q,\widetilde{Q}$ have dimension
	\ba{\bs{
\Delta[Q^{(s)(t)(u)}]=\Delta[\widetilde{Q}^{(s)(t)(u)}]&= \frac{1}{2} \left( (1-\epsilon)(N-1) -2 (s+t+u) (1+\epsilon) \right),\\ -\frac{N-1}{2} &\leq s,t,u \leq \frac{N-1}{2}.
	}}
We generically find that some operators decouple, and at finite-$N$ there will be field theory corrections to the central charges \eqref{eq:asphere}, and thus corrections to the BBBW results. We will not discuss these corrections in general, and instead briefly point out some features in the $|z|=2,3$ cases.

Our construction does not reach $|z|=0,1$. The case of $|z|=2$, i.e. $m=0$, is studied field-theoretically in \cite{Maruyoshi:2016aim}, where they find that the central charges violate the Hofman-Maldacena bounds due to accidental symmetries from the decoupled operators along the flow to the IR SCFT.

For $z=3$, i.e. $m=1$, and $N>2$, the operator $\tr \mu^2$ decouples. $Q,\widetilde{Q}$ operators with
	\ba{
	| s+t+u| \geq \frac{-21+N(4N(2N-3)-11) + (N-1) \sqrt{ 9+4N(1+N)(9+7N(1+N))}}{2 (9+10N(1+N) +  \sqrt{ 9+4N(1+N)(9+7N(1+N))}}
	}
violate the unitarity bound. For instance, for $N=2$ and $z=3$, the three operators $Q^{(s)(t)(u)}$ with $s+t+u=1/2$, and one with $3/2$ would have R-charges that unitarity. In general, the decoupling of these operators could lead to violations of the Hofman-Maldacena bounds, but we do not pursue this direction here\footnote{A discussion of the chiral operators in the ring of these theories on a sphere and their decoupling is given in \cite{Fazzi:2016eec}, as well as some discussion on which geometries do not flow to SCFTs in the IR.}.

\section{General $C_{g,n}$ from $T_N^{(m_i)}$ building blocks}

\label{sec:cgn}

\subsection{Computing $a_{trial}$ and $c_{trial}$ for $g\neq 1$}
\label{sec:trialgen}

Now we consider the most general case of constructing a genus $g$ surface with $n$ maximal punctures from $T_N^{(m_i)}$ building blocks, and computing $a_{trial}$ and $c_{trial}$. One useful way to arrange this computation is to glue in chains of $T_N^{(m_i)}$ theories to the $n=0$ cases we constructed in Section \ref{sec:b3wanalogue}. The result will be a genus $g$ surface where the number of punctures depends on how many chains we add. Then, we can close arbitrarily many of these punctures via the Higgsing procedure discussed in Section \ref{sec:spheresec}. 

Let us begin with the class of theories we considered in Section \ref{sec:b3wanalogue}: start with $\ell_1$ $T_N^{(m_i)}(\sigma_i=-1)$ blocks, and $\ell_2$ $T_{N}^{(m_i)}(\sigma_i=+1)$ blocks, glued with $n_1$ $\mc{N}=1$ vectors, and $n_2$ $\mc{N}=2$ vectors, where $\ell_1+\ell_2=2g-2$, $n_1+n_2=3(g-1)$, and $n_1=3\ell_1$. 

\begin{figure}[t!]
\centering
\begin{subfigure}{.33\textwidth}
  \centering
  \includegraphics[width=0.9\linewidth]{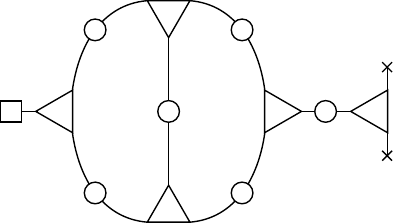}
  \caption{\ \ \ \ \ }
  \label{fig:sub1}
\end{subfigure}%
\begin{subfigure}{.33\textwidth}
  \centering
  \includegraphics[width=0.9\linewidth]{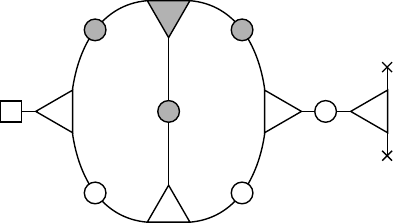}
  \caption{\ \ \ \ \ }
  \label{fig:sub2}
\end{subfigure}%
\begin{subfigure}{.33\textwidth}
  \centering
  \includegraphics[width=0.9\linewidth]{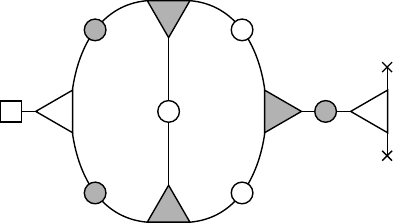}
  \caption{\ \ \ \ \ }
  \label{fig:sub3}
\end{subfigure}%
\caption{A genus 2, single-puncture example of a possible generalization of the diagrams in Figure \ref{fig:genus2ex}. In our notation, these all have $n_{tot}=1$, $n_{dif}=-1$, and $\ell_1+\ell_2=2,n_1+n_2= \ell_1'+\ell_2'= n_1'+n_2'=3,h_2=2$. All three diagrams have the same IR central charges.}
\label{fig:tests}
\end{figure}

\begin{figure}[t]
	\centering
	\includegraphics[width=0.7\textwidth]{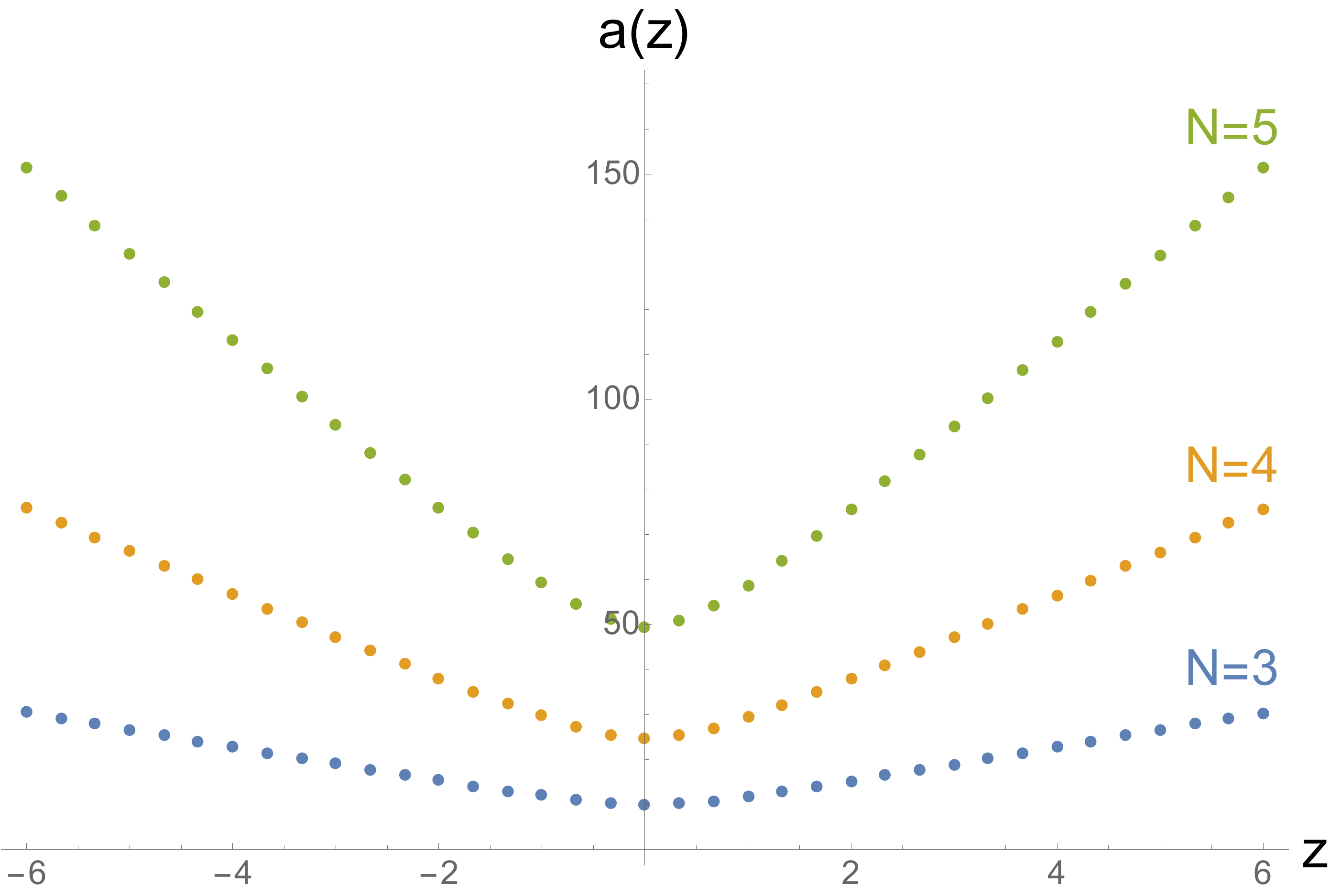}
	\caption{$a$ as a function of the twist $z$ for the $g=2,n_{tot}=1$ theories represented in Figure \ref{fig:tests}, plotted for $N=3,4,5$. \label{fig:apic0}}
	\end{figure}

Next, glue in some number of  additional $T_N^{(m_i)}$ blocks such that we do not change the genus of the surface. In particular, introduce $\ell_1'$ $T_N^{(m_i)}(\sigma_i=-1)$ blocks and $\ell_2'$ $T_N^{(m_i)}(\sigma_i=+1)$ blocks. This will require that we introduce $\ell_1'+\ell_2'$ vectors to glue in these chains, where the number of additional $\mc{N}=2$ vectors $n_2'$ is given by $\ell_2'-\ell_1'$. These $\ell_1'+\ell_2'$ blocks also introduce $\ell_1'+\ell_2'$ punctures to the surface, of various colors. As in Section \ref{sec:spheresec}, we can close each of these punctures by flipping their color, coupling in an adjoint chiral field, and letting the adjoint take a nilpotent expectation value that breaks the $SU(N)$ flavor group of the puncture. In particular, let us close $h_1$ of the $\sigma_i=-1$ punctures, and $h_2$ of the $\sigma_i=+1$ punctures. This will leave us with $n^{(-)} = \ell_1' - h_1$ minus punctures, and $n^{(+)} = \ell_2' - h_2$ plus punctures. An example of this construction applied to a genus $2$ surface that results in one plus puncture is given in Figure \ref{fig:tests}.

The total degrees $p$ and $q$ for the surface will be given by
	\ba{\bs{
	 p &= \sum_{\{\sigma_i=+1\} } m_i -  \sum_{\{\sigma_i=-1\} } m_i+ (\ell_2+\ell_2') - h_1, \\
	 q &= - \sum_{\{\sigma_i=+1\} } m_i + \sum_{\{\sigma_i=-1\} } m_i+ (\ell_1+\ell_1') - h_2,\\
	 p+q &= 2g-2 + n^{(-)} + n^{(+)} =-\chi .
	 } \label{eq:pqs}
	}
The sum over $\{\sigma_i=+1\}$ runs over the $\ell_2+\ell_2'$ plus nodes, while the sum over $\{\sigma_i=-1\}$ runs over the $\ell_1+\ell_1'$ minus nodes. Below, we write out answers in terms of $z=(p-q)/(p+q)$, the combinations 
	\ba{
	n_{tot} \equiv n^{(-)} + n^{(+)},\quad n_{dif} \equiv n^{(-)} - n^{(+)},
	}
and the Euler characteristic of the surface, $\chi=-2g+2-n_{tot}$. Summing the contributions to the trial central charges, we find $a(\epsilon)$ is given by
\ba{
\bs{
a(\epsilon)&=-\frac{3}{64}(N-1) \bigg[(1+N)  \left( 2 n_{tot} +  n_{dif}\epsilon(1-3\epsilon^2)\right)\\
&  + \chi \left( 2+3(1-\epsilon^2)N(1+N) -  z \epsilon \left( 1+3N(1+N) + 3\epsilon^2(1+N+N^2)\right) \right) \bigg],
}\label{eq:agens}
}
and $c(\epsilon)$ is given by
\ba{ \bs{
c(\epsilon)&=-\frac{1}{64}(N-1) \bigg[ -4 - (1+N)(9N-5\epsilon + 9n_{dif}\epsilon^3) + n_{tot} \left( 1-N(5+9N)\right)  \\
&+\chi\left(1-9\epsilon^2 N(1+N) + z\epsilon \left(-5+9\epsilon^2-9N(1+N)(1-\epsilon^2) \right) \right) \bigg].
}\label{eq:cgens}}
Maximizing with respect to $\epsilon$, we find that $\epsilon$ is given by
		\ba{\bs{
	\epsilon =& \frac{1}{3  \left( \chi z (1+N+N^2) - n_{dif}(1+N) \right) }\cdot   \\
	&\  \Bigg[\chi N (1+N)+ \bigg( n_{dif}(1+N)\left( (1+N)(n_{dif}-4Nz\chi) -2\chi z \right)  \\
	&\ \  +\chi^2 \left( N^2 (1+N)^2 + z^2(1+N(1+N)(4+3N(1+N) )) \right)  \bigg)^{1/2}\Bigg].
	} \label{eq:epsfinal}}
For $n_{tot}=n_{dif}$=0, these formulae reproduce \eqref{eq:a0} and \eqref{eq:c0}; for $g=n_{tot}=n_{dif}=0$ they reproduce \eqref{eq:asphere}.

At large-$N$, $\epsilon$ computed in \eqref{eq:epsfinal} matches the BBBW result listed in \eqref{eq:bbbwlargen}. Then, the leading order piece of $a$ and $c$ in a large-$N$ expansion is given by
	\ba{
	a_{\text{large-}N}=\frac{ N^3(2g-2+n_{tot} ) \left(9z^2-1 + (1+3z^2)^{3/2} \right)}{96z^2}.
	}
This is simply our result \eqref{eq:alargen} with $(2g-2)\to (2g-2+n_{tot})$. Thus at large $N$, the central charges depend on the Riemann surface only through the Euler characteristic $\chi=-2g+2-n_{tot}$.

We highlight that these answers depend only on topological data: the total number of punctures $n_{tot}$, the difference between the number of plus and minus punctures $n_{dif}$, the twist $z$ defined in terms of the normal line bundle degrees $p$ and $q$, and the genus $g$ of the surface (or equivalently, the Euler characteristic). It is satisfying that the computation organized such that the other parameters dropped out, leaving the (relatively) nice expressions \eqref{eq:agens}-\eqref{eq:epsfinal}.

\subsection{Comments on operators}

Deferring a detailed analysis of the chiral operators of this class of theories to the future, we point out one interesting feature in the chain operators $\mc{O}_{M2},\ \widetilde{O}_{M2}$. For illustration, consider the theories in \ref{fig:tests}. There will be $N^2$ fundamentals (and anti-fundamental) operators of the form $(QQQQQ)_\ell^{(s)(t)}$, where $-\frac{(N-1)}{2} \leq s,t \leq \frac{(N-1)}{2}$ label the $N^2$ operators and yield a shift in the $J_-$ charges of these operators as in Table \ref{tab:tnmhiggsedmatter}, and $\ell$ runs from $1,\dots,N$ is an index for the $SU(N)$ flavor symmetry of the remaining puncture. If we close the remaining puncture, then there will be $N^3$ degenerate chain operators that should correspond to M2-branes wrapping the surface. 

More generally, for a class of theories constructed from gluing $h_1$ Higgsed shaded flavor groups and $h_2$ Higgsed unshaded flavor groups such that all the punctures are closed, there will be gauge-invariant chain operators
	\ba{
	\mc{O}^{(s_1)\dots (s_{h_1}) (t_1)\dots (t_{h_2})} =\prod_{i=1}^{2g-2 + h_1+h_2} (Q_i)^{ (s_1)\dots (s_{h_1} ) (t_1)\dots (t_{h_2})},\quad -\frac{(N-1)}{2} \leq s_i,t_i \leq \frac{N-1}{2}
	}
as well as the corresponding operators constructed from the $\widetilde{Q}$'s. The R-charges of these operators will be given by
	\ba{\bs{
R\left(\mc{O}^{(s_1)\dots (s_{h_1}) (t_1)\dots (t_{h_2})} \right) =& \frac{1}{2} \Bigg(  \left((N-1)(h_2+\ell_2) - 2h_1 \sum_{i=1}^{h_1} s_{i} \right) (1-\epsilon) \\
&+ \left( (N-1) (2g-2+h_1-\ell_1) - 2h_2 \sum_{i=1}^{h_2} t_{i} \right) (1+\epsilon) \Bigg).
	}}
Thus, this field-theoretic analysis suggests a degeneracy of possible M2-brane operators\footnote{This degeneracy of operators was noted independently in  \cite{Fazzi:2016eec}.}.

\subsection{Computing $a_{trial}$ and $c_{trial}$ for the torus}\label{sec:torus} 

Up to this point, we've considered $g\neq 1$. The case of M5-branes compactified on the 2-torus is special because the torus admits a flat metric, implying that the maximal amount of supersymmetry can be preserved by fixing the normal bundle to the M5-brane worldvolume to be trivial. The singular behavior at $g=1$ in the computation of $a$ and $c$ is related to the fact that the M5-brane tension causes the volume of the torus to shrink. In our constructions, this means that we should get a 4d $\mc{N}=4$ field theory in the IR when $g=1$ and $z=0$.

One can formulate a nonsingular construction that preserves only $\mc{N}=1$ supersymmetry by taking the torus to have line bundles of equal and opposite degrees fibered over it, i.e. taking $p=-q$. Letting $g=1$ and $p=-q$ in \eqref{eq:pqs} means that we should require the total number of punctures be zero. In the construction detailed in Section \ref{sec:trialgen}, the simplest generalized quiver that this could correspond to is closing one of the punctures on a $T_N^{(m)}$ block, and then gluing the other two.

Redoing the computation of Section \ref{sec:trialgen} for the torus, we find
	\ba{
g=1:\quad	\epsilon = -\frac{1}{3} \sqrt{ \frac{1+3N(1+N)}{1+N+N^2}},
	}
and
	\ba{
g=1:\quad	a = \frac{p(N-1) (1+3N(1+N))^{3/2}}{48\sqrt{1+N+N^2}}. \label{eq:atorus}
	}
Indeed, \eqref{eq:atorus} matches the BBBW result that we've written in \eqref{eq:bbbwtorus}, where our definition of $p$ matches their $|z|$.

\section{Future directions} \label{sec:discussion}

There are many directions one can think about based on the present work, some of which we will list below.

First, we note that the authors of  \cite{Fazzi:2016eec} study the chiral rings of these theories (much as  \cite{Maruyoshi:2013hja} and \cite{Hayashi:2014hfa} studied the chiral rings of the $T_N$ theory in detail). It would be further interesting to study the moduli space of vacua of the 4d $\mc{N}=1$ SCFTs realized by our constructions.

In the present work we don't discuss possible confinement of the gauge theories that result from the gluing procedure. However, as shown in \cite{Agarwal:2015vla}, when two $T_N^{(m)}$ blocks with the same $m$ and opposite shading are glued with an $\mc{N}=1$ vector multiplet, the gauge node confines. It would be interesting to study the structure of confinement for our general constructions, as well as to understand how various duality maps of $T_N^{(m)}$ theories (discussed in \cite{Agarwal:2015vla}, and \cite{Fazzi:2016eec}) act on our constructions.

It would be interesting to obtain a field-theoretic construction of the $N=2,\ g=n=0$ theories with $|z|=0,1$. While these theories do not have a conformal phase, it could be useful to study the IR dynamics from the field theory side\footnote{The author is grateful to Ibrahima Bah for pointing out this possibility.}.

Also, it is shown in \cite{Bah:2013wda} that the BBBW solutions with rational central charges allow probe M5-branes to break into multiple M5-branes at special points; in particular, when $a$ and $c$ at large-$N$ (given in \eqref{eq:alargen}) are rational. It would be interesting to understand field-theoretically what happens at these special points in our constructions.


\section*{\centering Acknowledgements}


It is a great pleasure to thank Ibrahima Bah for suggesting the idea for this work, and for many  subsequent helpful conversations and comments. We are especially grateful as well to Ken Intriligator for many informative and valuable conversations. We thank Prarit Agarwal and Jaewon Song for helpful suggestions and comments pertaining to this work. This work was supported in part by DOE grant DE-SC0009919.

\appendix

\section{Conventions and main $T_N$ formulae} \label{sec:conventions} 

The $T_N^{(m)}$ theory is $\mc{N}=2$ supersymmetric with global symmetries $SU(2)_R\times U(1)_R\times SU(N)^3$. We use a basis for the Cartan subalgebra of the $\mc{N}=2$ R-symmetry $SU(2)_R\times U(1)_R$ labeled by $(I_3,R_{\mc{N}=2})$. The R-symmetry of an $\mc{N}=1$ subalgebra is given by
	\ba{
	R_{\mc{N}=1}=\frac{1}{3} R_{\mc{N}=2} + \frac{4}{3}I_3.
	}
We can rewrite these in terms of the generators $J_+,J_-$ of the $U(1)_+\times U(1)_-$ symmetry preserved by the $\mc{N}=1$ theories of class $\mc{S}$, using $R_{\mc{N}=2}=J_-$, and $I_3=\frac{1}{2} J_+$. With these conventions, for example, the adjoint field in the $\mc{N}=2$ vector multiplet has $(J_+,J_-)=(0,2)$.

The $(J_+,J_-)$ charges of chiral operators of the $T_N$ theory are
		\begin{center}
	\begin{tabular}{c|c}
	 & $(J_+,J_-)$ \\ \hline
  $u_k$ &  $(0,2k)$ \\
   $Q,\widetilde{Q}$ &  $(N-1,0)$ \\
    $\mu_{A,B,C}$ &  $(2,0)$ 
    \end{tabular} \label{fig:charges}
    \end{center}	
The $\mu_{A,B,C}$ are moment-map operators in the adjoint of (one of) the $SU(N)_A\times SU(N)_B\times SU(N)_C$ flavor symmetry groups,  and the $Q$($\widetilde{Q}$) transform in the trifundamental(anti-trifundamental) of the $SU(N)_A\times SU(N)_B\times SU(N)_C$ symmetry.  The $u_k$ are Coulomb branch operators of dimension $k$, with $k=3,\dots,N$.

The IR superconformal R-charge for operators of the $T_N$ theory of color $\sigma_i=\pm1$ is given by maximizing the following combination of R-charges with respect to $\epsilon$:
	\ba{ \bs{
	R(\epsilon) &= \left(\frac{1}{2} R_{\mc{N}=2} + I_3\right) + \sigma_i\epsilon \left(\frac{1}{2} R_{\mc{N}=2} - I_3 \right)\\
	&= \frac{1}{2} (1-\sigma_i \epsilon)J_+ + \frac{1}{2} (1+\sigma_i \epsilon) J_- .}\label{eq:tnrcharge}
	}

\section{Relevant BBBW results} \label{sec:bbbwapp} 

In \cite{Bah:2012dg}, Bah, Beem, Bobev, and Wecht (BBBW) compute $a$ and $c$ of the IR $\mc{N}=1$ SCFTs obtained from compactifying the 6d (2,0) theories on a Riemann surface $\mc{C}_g$, where the surface is embedded in a Calabi-Yau three-fold that decomposes into a sum of line bundles as in \eqref{eq:cygeometry}. These are computed by integrating the anomaly eight-form of the M5-brane theory over the surface $\mc{C}_{g}$, and matching with the anomaly six-form, which is related to the anomalous divergence of the 4d $\mc{N}=1$ R-current by the descent procedure and given by
	\ba{
	I_6=\frac{1}{6} \tr R^3 c_1(F)^3 - \frac{1}{24} \tr R c_1(F) p_1(T_4).
	}
$F$ is the $S^1$ bundle which couples to the R-symmetry, $T_4$ is the tangent bundle to the 4d spacetime manifold, $c_1$ is the first Chern class, and $p_1$ is the first Pontryagin class. Due to the presence of an additional global symmetry $U(1)_{\mc{F}}$, the superconformal R-symmetry takes the form
	\ba{
	R = K +\epsilon \mc{F},
	}
where $\epsilon$ is a real number determined by $a$-maximization. For the (2,0) theory of type $A_{N-1}$, $\epsilon$ is found to be 
	\ba{
\epsilon = \frac{\eta + \kappa \zeta}{3(1+\eta)z},	 
	}
and the central charges $a$ and $c$ are found (for $g\neq 1$) to be
	\ba{ \bs{
	a&=(g-1) (N-1) \frac{\zeta^3 + \kappa \eta^3 - \kappa (1+\eta) (9+21\eta + 9 \eta^2)z^2}{48(1+\eta)^2 z^2},\\
	c&=(g-1)(N-1)  \frac{\zeta^3 + \kappa \eta^3 - \kappa (1+\eta) (6-\kappa \zeta + 17\eta + 9 \eta^2)z^2}{48(1+\eta)^2 z^2}.
	} \label{eq:abbbw}}
$\eta$ and $\zeta$ are defined as
	\ba{
	\eta = N(1+N),\quad \zeta = \sqrt{\eta^2+(1+4\eta + 3 \eta^2)z^2}.
	}
$z$ is the twist parameter defined in terms of the degrees of the line bundles $p$ and $q$ as in \eqref{eq:zdef}, and $\kappa=1$ for the sphere and $\kappa=-1$ for a hyperbolic Riemann surface. In the large $N$ limit, these simplify to
	\ba{ \bs{
	\epsilon_{\text{large-}N} &= \frac{1+\kappa \sqrt{1+3z^2}}{3z}\\
	a_{\text{large-}N} &=c_{\text{large-}N} = (1-g) N^3 \left(\frac{1-9z^2+\kappa (1+3z^2)^{3/2}}{48 z^2} \right).
	} \label{eq:bbbwlargen}}
The computation for $g=1$ requires special care, as one can preserve $\mc{N}=4$ supersymmetry in the IR by fixing the normal bundle to the M5-brane worldvolume theory to be trivial. However, taking $p=-q$ preserves only $\mc{N}=1$ supersymmetry in the IR. Redoing the computation for this special value, BBBW find that for the $A_{N-1}$ theory on the torus,
	\ba{ \bs{
	\epsilon &= -\frac{1}{3} \sqrt{\frac{1+3\eta}{1+\eta}},\\
	a&= \frac{|z|}{48} \frac{(N-1)(1+3\eta)^{3/2}}{\sqrt{1+\eta}},\quad c=\frac{|z|}{48} \frac{(N-1) (2+3\eta)\sqrt{1+3\eta}}{\sqrt{1+\eta}},} \label{eq:bbbwtorus}
	}
where at large-$N$,
	\ba{
	a_{\text{large-}N}=c_{\text{large-}N} = \frac{\sqrt{3}}{16} |z| N^3.
	}

\section{'t Hooft anomalies for gluing $T_N^{(m)}$ building blocks} \label{sec:anomalies} 

The 't Hooft anomaly coefficients for a single $T_N$ block are given by 
\ba{ 
\begin{array}{c|c}
	& T_N\ \text{coefficients} \\ \hline
	J_+,J_+^3 & 0 \\
	J_-,J_-^3 & (N-1) (-3N-2)  \\
	J_+^2J_- & \frac{1}{3} (N-1) (4N^2-5N-6) \\
	J_+J_-^2 & 0 \\
	J_+SU(N)^2_{A,B,C} & 0 \\
	J_-SU(N)^2_{A,B,C} & -N 
\end{array}
}
An $\mc{N}=2$ vector multiplet contains two fermions with $(J_+,J_-)=(1,1)$ and $(-1,1)$, so the only nonzero anomaly coefficients are 
	\ba{
	\mc{N}=2\ \text{vector}:\quad J_-=J_-^3=J_+^2J_- = 2(N^2-1).
	}
Consider Higgsing an $SU(N)$ flavor group on a $T_N$ block by giving a nilpotent vev to the adjoint chiral multiplet, $\langle M \rangle = \rho(\sigma_3)$, where the $SU(N)$ flavor corresponds to a maximal puncture whose color is opposite the background color. This can be computed\footnote{The author is grateful to Prarit Agarwal for explaining this computation in more detail.} by shifting $J_-\to J_- - 2\rho(\sigma_3)$ and summing the contribution from the remaining $N-1$ singlets $M_j,\ j=1,\dots,N-1$ whose R-charges are shifted to $(J_+,J_-)=(0,2+2j)$. This results in the following contribution to the block being Higgsed:	
	\ba{
	\begin{array}{c|c}
	&  \text{from Higgsing}\\ \hline
	J_+,J_+^3 & 1-N \\
	J_-,J_-^3 & N^2-1  \\
	J_+^2J_- & N^2-1 \\
	J_+J_-^2 & \frac{1}{3}(1-N)(4N^2+4N+3) \\
	J_+SU(N)^2_{A,B,C} & 0 \\
	J_-SU(N)^2_{A,B,C} & 0 
	\end{array} \label{eq:higgsedanom}}
The 't Hooft anomaly coefficients for a single $T_N^{(m)}$ block are computed in \cite{Agarwal:2015vla} by summing the contributions of $(m+1)$ $T_N$ blocks---$m$ of which have a Higgsed flavor group---and $m$ $\mc{N}=2$ vector multiplets, yielding 
	\ba{
	\begin{array}{c|c}
	& T_N^{(m)}\ \text{coefficients} \\ \hline
	J_+,J_+^3 & m(1-N) \\
	J_-,J_-^3 & (N-1) (m-3N-2)  \\
	J_+^2J_- & \frac{1}{3} (N-1) (4N^2-5N-6+m(4N^2+4N+3)) \\
	J_+J_-^2 & \frac{1}{3} m (3+N-4N^3) \\
	J_+SU(N)^2_{A,B,C} & 0 \\
	J_-SU(N)^2_{A,B,C} & -N 
	\end{array} \label{eq:tnmanomalies}
	}
Taking $m=0$ reproduces the $T_N$ 't Hooft anomalies. 

Given these anomaly coefficients, we can compute the contribution to the central charges, using
	\ba{ \bs{
	&a(\epsilon)=\frac{3}{32}\left( 3 \tr R(\epsilon)^3 - \tr R(\epsilon) \right)\\
	&= \frac{3}{64} \bigg( \frac{3}{4} \bigg[ (1-\epsilon)^3 J_+^3 + (1+\epsilon)^3 J_-^3 + 3(1-\epsilon)^2 (1+\epsilon) J_+^2J_- + 3(1-\epsilon)(1+\epsilon)^2 J_+J_-^2 \bigg] \\
	&\quad\quad\ \ \   -(1-\epsilon)J_+ - (1+\epsilon)J_-  \bigg).
	}}

\bibliography{bibfile}{}
\bibliographystyle{utphys}

\end{document}